\documentclass[doublecol]{epl2}
\pdfoutput=1

\newcommand{\widthHalfColumn}{4.25cm}
\newcommand{\widthFullColumn}{8.5cm}
\newcommand{\widthTwoColumns}{17cm}

\usepackage{graphics}
\usepackage{graphicx}
\usepackage{epsfig}
\usepackage{amssymb}
\usepackage{amstext}
\usepackage{bm}
\usepackage{times}

\renewcommand{\vec}[1]{\mathbf{#1}}
\newcommand{\ket}[1]{\ensuremath{|#1\rangle}}
\newcommand{\expect}[1]{\ensuremath{\langle #1 \rangle}}
\newcommand{\ER}{\ensuremath{E_R}}
\newcommand{\Vh}{V_h}

\newcommand{\AFFITP}{{I. Institut f\"ur Theoretische Physik, Universit\"at Hamburg,
  Jungiusstra{\ss}e 9, 20355 Hamburg, Germany} }
\newcommand{\AFFMCUA}{{MUARC, School of Physics and Astronomy,
University of Birmingham, Edgbaston, Birmingham B15 2TT, UK} }

\renewcommand{\section}[1]{~\\[-0.5ex]\noindent{\bf #1}\\[1ex]}

\begin{document}

\title{\rm \bf \LARGE  Excitation spectrum of Mott shells in optical lattices}
\shorttitle{Excitation spectrum of Mott shells in optical lattices}

\author{\bf Dirk-S\"oren L\"uhmann\inst{1}\hspace{-1ex},\hspace{0.5ex} Kai Bongs\inst{2}\hspace{-1ex} and\hspace{0.2ex} Daniela Pfannkuche\inst{1}}
\institute{
\inst{1} \rm\AFFITP\\
\inst{2} \rm\AFFMCUA \vspace{-3.5ex}
}
\shortauthor{D.-S. L\"uhmann \etal}
\pacs{67.85.Hj}{Bose-Einstein condensates in
optical potentials\vspace{2.0ex}}

\abstract{\raggedright
We theoretically study the excitation spectrum of confined 
macroscopic optical lattices in the Mott-insulating limit. 
For large systems, a fast numerical method is proposed to 
calculate the ground state filling and excitation energies.
We introduce many-particle on-site energies capturing multi-band effects
and discuss tunnelling on a perturbative level using an effectively 
restricted Hilbert space.   
Results for small one-dimensional lattices obtained by this method
are in good agreement with the exact multi-band diagonalization of the
Hamiltonian.
Spectral properties associated with the formation of regions
with constant filling, so-called Mott shells, are investigated
and interfaces between the shells with strong particle fluctuations are 
characterized by gapless local excitations. 
}

\topmargin-2.02cm
\textheight24.5cm
\maketitle

\section{Introduction} 
The equivalence of lattice sites is the foundation of solid state physics
as it causes bands and lattice-periodic wave functions.
Interesting physics emerges when the 
band is bend caused by the inhomogeneity of the system,   
e.g. in clusters, where  bulk and surface  electrons show a different physical behaviour.
In optical lattices, the inhomogeneity comes  naturally due to  the finite waist of
the lattice-establishing laser beams or  due to  an additional dipole trap.
It has been shown in a series of pioneering work \cite{Jaksch,Greiner:415,stoferle:130403} that the 
bosonic Mott-insulator phase can be realized in optical lattices. 
Further measurements \cite{folling:060403,Campbell:313:649,cheinet:090404} 
have demonstrated that the situation is more subtle and 
Mott-insulator shells appear, i.e. plateaus with constant filling factors
descending in integer steps from the centre of the trap.
The existence of superfluid regions between the Mott shells  
has been theoretically predicted 
\cite{Batrouni:117203,Kashurnikov.66.031601,wessel:053615,Kollath.69.031601,demarco:063601,gerbier:053606}.  
The system has previously been studied numerically, using quantum Monte Carlo \cite{Batrouni:117203,Kashurnikov.66.031601,wessel:053615} and DMRG \cite{Kollath.69.031601},
and analytically in Refs.~\cite{demarco:063601,gerbier:053606}.  
In addition to condensed matter aspects, the coexistence of compressible and incompressible regions
has important implications on adiabatic heating in optical lattices \cite{ho:120404}.

We present a multi-band exact diagonalization study of small
systems exploring the exact excitation spectrum and the
precursor of shell formation. 
Based on this calculation  
we use a numerical method, which 
in the limit of deep lattices 
allows us to obtain an approximated energy spectrum and
the exact occupation numbers.
Thereby, we incorporate orbital changes by    
introducing a particle-number-dependent on-site interaction.  
The formation of shells, local excitation gaps and particle fluctuations are discussed
reflecting the strong inhomogeneity of the system.
Our approach is suitable for optical lattices with millions of atoms
in arbitrary spatial dimensions and allows a perturbative treatment of
tunnelling. 
Compared to algorithms such as quantum Monte Carlo \cite{Batrouni:117203,Kashurnikov.66.031601,wessel:053615} and
DMRG \cite{Kollath.69.031601}, this technique gives numerically inexpensive 
results for large 3D lattice systems in a specific parameter regime.  
We organize this paper as follows. 
Subsequent to the introduction of the system, we present results
obtained by means of exact diagonalization followed by 
the numerical approach for vanishing tunnelling and
a comparison of both results. 
Finally, we discuss consequences for two- and three-dimensional optical lattices
including non-zero tunnelling.

The interaction of the ultracold bosonic atoms with mass $m$ is modelled
by a contact potential  $g\delta(\vec r - \vec r')$ with $g=\frac{4\pi \hbar^2}{m} a_s$ and the $s$-wave scattering length $a_s$.
Using the bosonic field operator  $\hat\psi(\vec r)$, the Hamiltonian including the repulsive two-particle interaction 
reads 
\begin{eqnarray}
  \hat H&=&\int d^3r \left( \hat\psi^\dagger(\vec r) \left[
     \frac{ {\vec p}^2}{2m}+ V_P(\vec r)+ V_{C}(\vec r)\right] \hat\psi(\vec r) \right.\nonumber\\
& &\left.+\frac{g}{2} \hat\psi^\dagger(\vec r) \hat\psi^\dagger(\vec r) \hat\psi(\vec r)
  \hat\psi(\vec r)  \right),
  \label{H}
\end{eqnarray}
where the periodic potential $V_P$ of the optical lattice is given by 
$V_{0,x}\cos^2(\pi x /a)+V_{0,y}\cos^2(\pi y /a)+V_{0,z}\cos^2(\pi z /a)$
with the lattice constant $a$.
The atoms experience an additional confinement potential $V_C$
caused by the finite Gaussian beam waist  $W_0$  
of the laser beams
and/or an additional dipole trap with the frequency $\omega_d$.
Using a harmonic approach, the confining potential in $x$-direction is given by
\begin{equation}
 V_{C}=\Vh\, \frac{x^2}{a^2}, 
\end{equation}
where $\Vh=\frac{1}{2}m\omega^2_\text{eff} a^2$ with
$\omega^2_\text{eff}\approx\frac{4(V_{0,y}+V_{0,z})}{m W_0^2}+\omega^{2}_{d,x}$.

\begin{figure*}[t]
\centering\includegraphics[width=\widthTwoColumns]{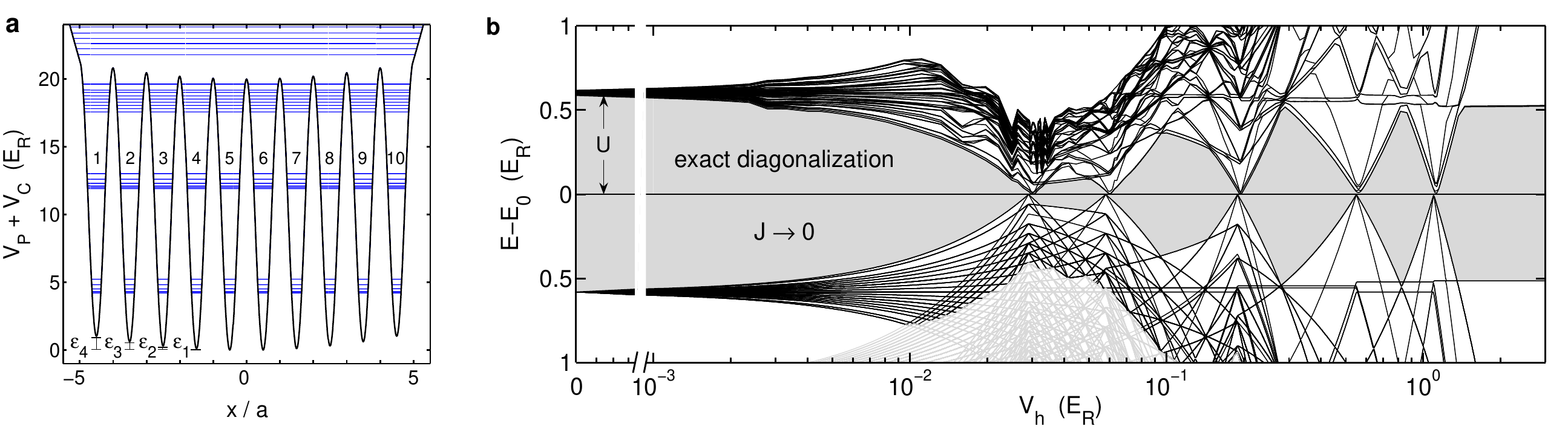}
\caption{(a) The periodic potential $V_P$ is superposed by
a harmonic potential $V_C$ leading to site offsets $\epsilon_i$.  
The single-particle spectrum  for $\Vh=0.06\ER$ 
containing discrete bands is plotted within the potential.
(b) The many-particle energy spectrum  ($91$ states)  obtained by exact diagonalization (upper half) 
and for vanishing tunnelling (lower half)
as a function of the harmonic confinement strength $\Vh$.  
Additional states in the $J\rightarrow0$ spectrum are plotted in grey.  
The energy scale is relative to the ground state energy $E_0$ and
the gap to the first excited state is depicted in grey.}
\label{spectrum}
\end{figure*}

\section{Exact diagonalization}
For the exact diagonalization, the potential is truncated to a 1D lattice with $m_x=10$ and $m_y=m_z=1$ sites 
by adding a smooth boundary \cite{luhmann:023620}.
Exemplarily, we have chosen $^{87}$Rb atoms with $a_s=100 a_0$, where $a_0$ is the Bohr radius,
$N=10$ particles, a lattice constant $a=515\text{~nm}$, and a transversal lattice depth $V_{0,y}=V_{0,z}=40\ER$, 
where $E_R=\frac{h^2}{8 m a^2 }$ is the recoil energy.
The lattice depth in the longitudinal direction is $V_{0,x}=20\ER$,
leading to a system deep within the Mott-insulator phase for a 
vanishing harmonic confinement ($\Vh=0$). 
The potential of the 1D lattice is shown in figure~\ref{spectrum}a, 
depicting the  energy offsets $\epsilon_i$ of the modeled chain,  
which are labeled $\epsilon_0=0$ for the two equivalent central sites
and $\epsilon_4$ for the outermost sites.  
The parabolic shape of the confinement leads to large energy offsets
of outer sites, although the offsets are small for central ones. 
At $\Vh=0$, the single-particle spectrum 
shows three discrete bands that  comprise ten delocalized Bloch states each  
(corresponding to the number of sites),
where the width of the lowest band is $4J\approx 2.4\times 10^{-3}\ER$ \cite{note1}.  
For larger values of $\Vh$  the `band' width is given basically by the offset of the outermost sites
(see figure~\ref{spectrum}a). 
At $\Vh \gtrsim 0.4\ER$, the  outermost  site  offset $\epsilon_4 \gtrsim 8\ER$ causes the lowest two
single-particle bands to overlap.  

To calculate the many-particle spectrum by means of exact diagonalization, we limit the many-particle basis
to  $500\,000$ Fock states with lowest energy  
and freeze out the orbital degrees of freedom in $y$- and $z$-direction.
Afterwards, we calculate the matrix elements of (\ref{H}) and 
obtain the $91$ lowest eigenvalues (the ground state and
the $90$ states of the first band for $V_h\rightarrow 0$).
The exact diagonalization method includes particle correlations and the admixture of higher bands, 
i.e., orbital changes taking place for higher fillings \cite{luhmann:023620},
but is strongly limited to systems with few lattice sites.
The many-particle spectrum is plotted in figure~\ref{spectrum}b (upper half)
relative to the ground state energy $E_0$ 
as a function of the harmonic confinement $\Vh$ using a logarithmic scale.
For convenience, the band gap between the ground state and the first
excited state is depicted in grey.

For vanishing confinement ($\Vh=0$), the spectrum shows that
the first excited band is gapped from the ground state by 
the on-site interaction energy of two particles $U\approx0.6\ER$,
which corresponds to the gap of the macroscopic Mott-insulator phase. 
The atoms are strongly localized on single lattice sites \cite{luhmann:023620},
so that all lattice sites are occupied by exactly one particle per site.
Increasing the harmonic confinement 
leads to abrupt crossovers to states with higher integer occupation numbers,
i.e. finite size correspondents to Mott shell configurations.
In the many-particle spectrum these crossover points are reflected
by vanishing energy gaps.
In between these points lobe-like energy gaps can be observed, where
the lobes correspond to the occupation number configurations 
1-1-1-2-2-1-1-1, 1-2-2-2-2-1, 2-3-3-2, 1-4-4-1, and finally for
$\Vh>1\ER$ a states with 5 particles on the two central sites \cite{note2}.
For $\Vh<0.03\ER$, the lattice is commensurately filled with one
particle per site.
In this region, the excited band broadens with increasing $\Vh$,  leading
to a slowly decreasing energy gap. 
At $\Vh\approx 0.03\ER$, where the energy gap vanishes,  
a multitude of low-lying excitations is possible leading
to a compressible system. 
At this point, the double occupation of the central sites, which  
corresponds to the on-site energy $U$,
becomes energetically preferable to the  occupation  
of the outermost sites  with $\epsilon_4=20\Vh$.  
Analogous situations exist at the other crossovers, so that 
the shell structure is dominated by the ratio of on-site interaction $\frac{n_i(n_i-1)}{2}U$ 
to individual site offsets $\epsilon_i$.  
The critical behaviour separating different filling configurations
manifests itself in the disappearance of the excitation gap.

\section{Classical approach  and orbital degrees of freedom }
The nature of the excited states, however, is more complicated and the energy spectrum,
which contains important physics, is rather complex. 
In the following, we therefore consider the `classical' case
of vanishing tunnelling ($J\rightarrow0$) which is valid deep within the Mott-insulator phase.
Using this approach, we show that the basic features of the spectrum can be uncovered.
Furthermore, larger systems can be studied, for which finite tunnelling can be reintroduced  
perturbatively  in a second step, dealing with a  
drastically  reduced basis set.  
For $J\rightarrow0$, the truncated Bose-Hubbard Hamiltonian is given by
\begin{equation}
	\hat{H}=\sum_i \frac{\hat{n}_i(\hat{n}_i-1)}{2} U + \epsilon_i \hat{n}_i
	\label{HU}
\end{equation}
and the localized occupation number basis $\ket{n_1,n_2,...,n_M}$ is
an eigenbasis of the Hamiltonian, where $M=m_x m_y m_z$ denotes the number of sites.
In principle, finding the ground state requires to calculate the total energy $E=\expect{\hat{H}}$
of all possible basis states. The efficiency of this method is very limited
due to the huge number of basis states $\frac{(N+M-1)!}{N!(M-1)!}$ for large lattices, where
$N$ is the number of particles. 
Following \cite{demarco:063601,gerbier:053606}, in the local density approximation an effective local chemical potential $\tilde{\mu}(\vec r)=\mu-V_{C}(\vec r)$ 
can be introduced, and the ground state  is then constructed via filling  
each lattice site up to the local chemical potential $\tilde{\mu}$ separately.   
Using the continuous limit, the chemical potential $\mu(N)$ is calculated analytically 
in \cite{demarco:063601} and numerically in \cite{gerbier:053606}.
The continuous limit is, however, only applicable to smooth confining potentials.  
In general, the self-consistent determination of $\mu(N)$ is tedious.
We therefore suggest the following iterative algorithm for the solution of (\ref{HU}) 
directly in the microcanonical ensemble with a fixed total particle number $N=\sum_i n_i$.
It allows us also to construct the lowest excited states of the system and therefore
a comparison with figure~\ref{spectrum}b (upper half).  
Starting with an empty lattice, the $N$ particles are added successively to  
that site $j$ of the lattice,  where the expense of energy  $\mu_j^+ = n_j U+\epsilon_j$ is presently minimal.  
This procedure gives the lowest energy occupation for any particle number $N$, 
since $\mu_j^+>0$ and all sites are uncorrelated.  
The complexity of this algorithm is given by $O(NM)$, since each step
adds one particle to one of $M$ possible sites. 
This allows the calculation of the  exact  occupation numbers for a million particles, 
e.g., $N=M=100^3$, within seconds on an ordinary desktop computer.
This method is therefore considerable useful for the  design and interpretation of experiments.

\begin{figure}[t]
\begin{center}
\includegraphics[width=\widthFullColumn]{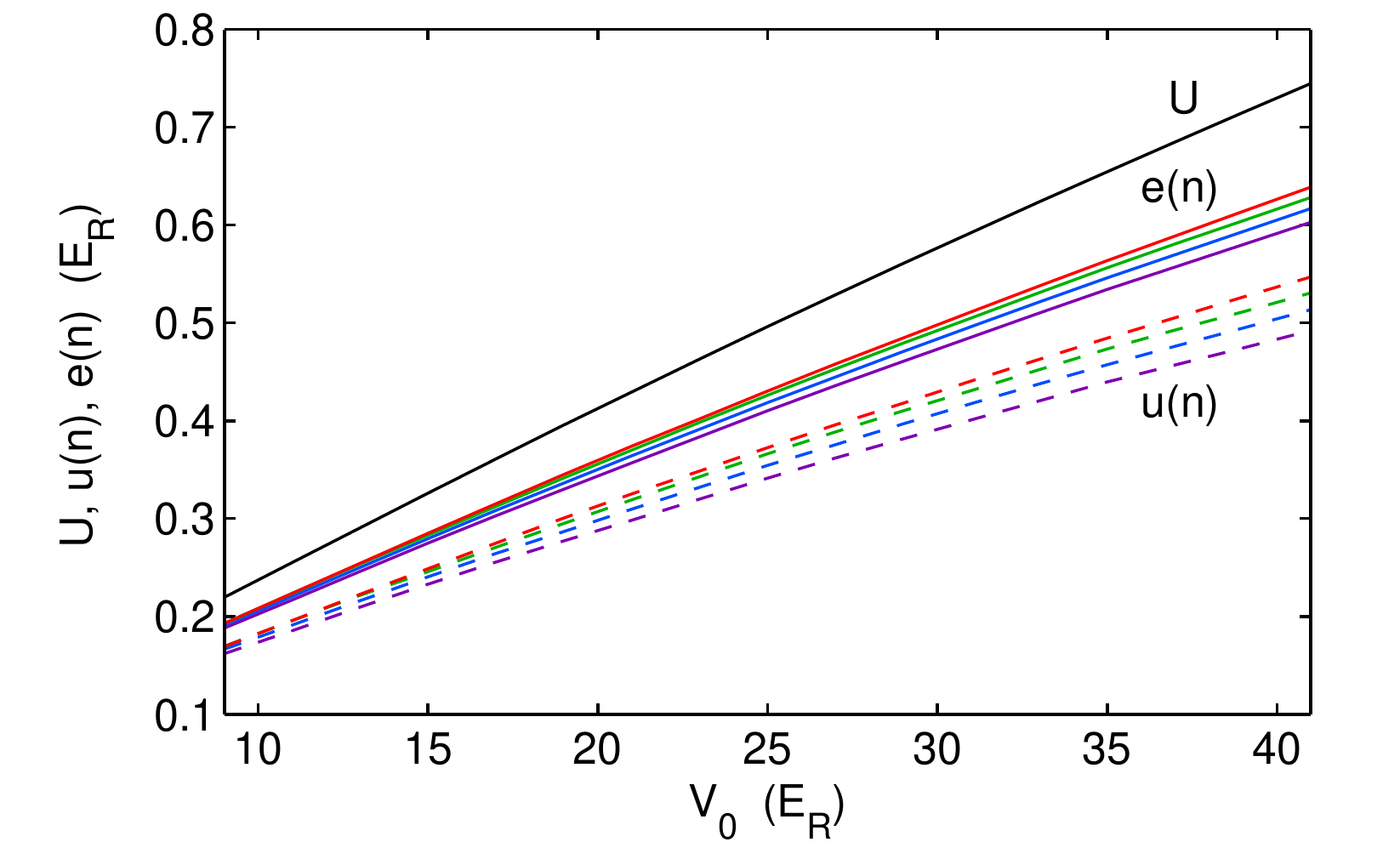}
\end{center}
\caption{A comparison of Hubbard on-site interaction $U$, the
many-particle interaction energy $u(n)$, and the total on-site energy $e(n)$ per atom pair
for $V_{0,i}=V_0$ and $i=x$, $y$, $z$.
The single-site many-particle wave function is determined by exact diagonalization
for $n=2$ particles (highest energy) to $n=5$  particles (lowest energy).  }
\label{multiorbit}
\end{figure}

Compared with the exact diagonalization results for the 1D lattice, the crossover points can, 
in principle, be reproduced with this method but their positions are shifted. This effect is   
additionally enhanced for higher filling factors.
To study this influence more quantitatively,    
we calculate the on-site few-particle interaction with exact diagonalization.
Since the lattice depth is sufficiently deep, it is valid to 
approximate the lowest-band Wannier function $w_0(\vec r)$
by the wave function of a single sinusoidal
lattice site using hard boundary conditions.
The correlationless Hubbard on-site interaction $U=g\int d^3r\, |w_0(\vec r)|^4$  
does not incorporate that particles tend to avoid each other for repulsive
interaction. This causes broadening of the particle density and was
addressed experimentally in \cite{Campbell:313:649}.
Using the correct many-particle wave function $\Psi_n$ for $n$ particles,
the expectation value of the on-site interaction becomes
\begin{equation}
	u(n)=\frac{g}{n(n-1)}\expect{\hat\psi^\dagger(\vec r) \hat\psi^\dagger(\vec r)
		 \hat\psi(\vec r)  \hat\psi(\vec r) },
\end{equation}
which is normalized to the interaction energy per atom pair.
The results that are depicted in figure~\ref{multiorbit} show
that $u(n)$ determined by exact diagonalization deviates
strongly from the Hubbard on-site interaction $U$.
As expected, the interaction energy $u(n)$ decreases with an increasing
number of particles per site.
However, not only $u(n)$ changes when the modification of wavefunctions
is taken into account. 
In fact, the admixture of correlated states and the broadening of the density 
change the expectation value of on-site kinetic and potential energy.
The total on-site energy is the eigenvalue of the many-particle
Schr\"odinger equation $\hat H_i\, \Psi_n=E_n\,\Psi_n$ restricted
to single-site wave functions and using the full Hamiltonian (\ref{H})
with $\Vh=0$.
The normalized on-site energies $e(n)=2(E_n-E_n^0)/n(n-1)$ are plotted in figure~\ref{multiorbit},
where $E_n^0$ is the energy of the non-interacting system. 
It shows, that the total interaction lies energetically between $U$ and $u(n)$.
Nevertheless, the deviation of $e(2)$ from the Hubbard $U$ is large. 
Using $U=e(2)$, the crossovers are still shifted noticeably for higher fillings
(large $\Vh$) comparing with the results for the 1D lattice 
(upper half of figure~\ref{spectrum}b).  
Therefore, we incorporate in our calculations 
the correct total  interaction  energy $e(n_j)$ for a single site with the occupation  
$n_j$. Please note that the   
optimization problem remains the same if substituting the energy  
$\mu_j^+ = n_j U+\epsilon_j$ for adding one particle by $\mu_j^+=e(n_j+1)-e(n_j)+\epsilon_j$.  
 In the lower half of figure~\ref{spectrum}b the energy spectrum for vanishing tunnelling ($J\rightarrow 0$)
is shown using the corrected values of the on-site interaction.  
In this case, the crossover energies corresponding to the vanishing gaps for $J\rightarrow 0$ are in good agreement with the exact diagonalization 
of the 1D lattice.   
This result shows, in general, that  the introduced filling-dependent on-site interaction  
$e(n)$  is appropriate to describe effects arising from orbital changes. 
The small remaining shift of the crossover energies  in figure~\ref{spectrum}b  is  due
to the  `classical'  treatment of the states in our approach.

The energy gap can be obtained
by removing one particle from site $j$ and adding it to site $k\neq j$.
Finding the minimum excitation energy 
$\Delta E_{j,k}$ for all possible $j$ and $k$ 
has in general the complexity $O(M^2)$.  
The excitation energy is given by   
$\Delta E_{j,k}=\mu_j^+ + \mu_k^- $, where
\begin{equation}
\mu_j^+ = e(n_j+1)-e(n_j)+\epsilon_j>0
\end{equation}
for adding a particle on site $j$ and 
\begin{equation}
\mu_k^- = e(n_k-1)-e(n_k)-\epsilon_k<0 
\end{equation}
for removing a particle from site $k$. Thus, 
it is sufficient to minimize 
$\mu_j^+$ and $\mu_k^-$ separately, which reduces the complexity to linear order in $M$.
Finding the next excited state is more complicated, since
this state may be an excitation of the ground state
but also of the first excited state.
For the calculation of the spectrum, a slightly 
modified Dijkstra algorithm can be used,
where one considers the ground state as the node of a graph.
This node is expanded according to all possible excitations. 
Iteratively, the node with the minimum energy, which is not expanded yet, 
is expanded. Consequently, the list of expanded nodes represents
the states with the lowest energy. 
It is necessary, however, to check whether a created excited state
is already a node in the graph, which is a major contribution 
to the complexity of the Dijkstra algorithm, 
but can numerically be highly optimized.  
Using this procedure,  the resulting energy spectrum (lower half of figure~\ref{spectrum}b)
reproduces well the basic features,
i.e. the band gap, the overall shape and the density of states,
of the exact calculation (upper half).
Because of interactions, the degeneracy of states
is lifted in the exact spectrum, so that the
actual energies of many states are shifted.

\section{Macroscopic lattices}
Transferring the above results to macroscopic lattices is not straightforward. 
Enlarging the system's size causes the differences in offset energies 
of neighbouring sites $\epsilon_{j}-\epsilon_{k}$
to decrease substantially when keeping the offset of the outermost sites fixed.
This causes the width and the height of the energy lobes in the spectrum to decrease  
because  a huge number of configurations become possible when increasing
the number of particles and lattice sites.
This process is drastically enhanced in 2D and 3D lattices, where  practically 
the band gap vanishes for all confinement strengths $\Vh$.
The only exception is the real Mott-insulator phase, where
the outermost sites have an offset smaller than $U$.
In fact, the excitation spectrum of the total system becomes more or less continuous.
This might appear contradictory to the insulating property at first glance
but the system is inhomogeneous and only some of the atoms can perform gapless excitations. 
Therefore, local properties are more suited to describe the system, such as
particle number fluctuations and the local excitation spectrum.
It has already been pointed out \cite{Kashurnikov.66.031601,Kollath.69.031601,wessel:053615,demarco:063601} 
that the regions  with fixed occupation number,
the Mott-insulator shells, are surrounded by compressible shells for 
non-negligible tunnelling.

\begin{figure}[t]
\begin{center}
\includegraphics[width=\widthFullColumn]{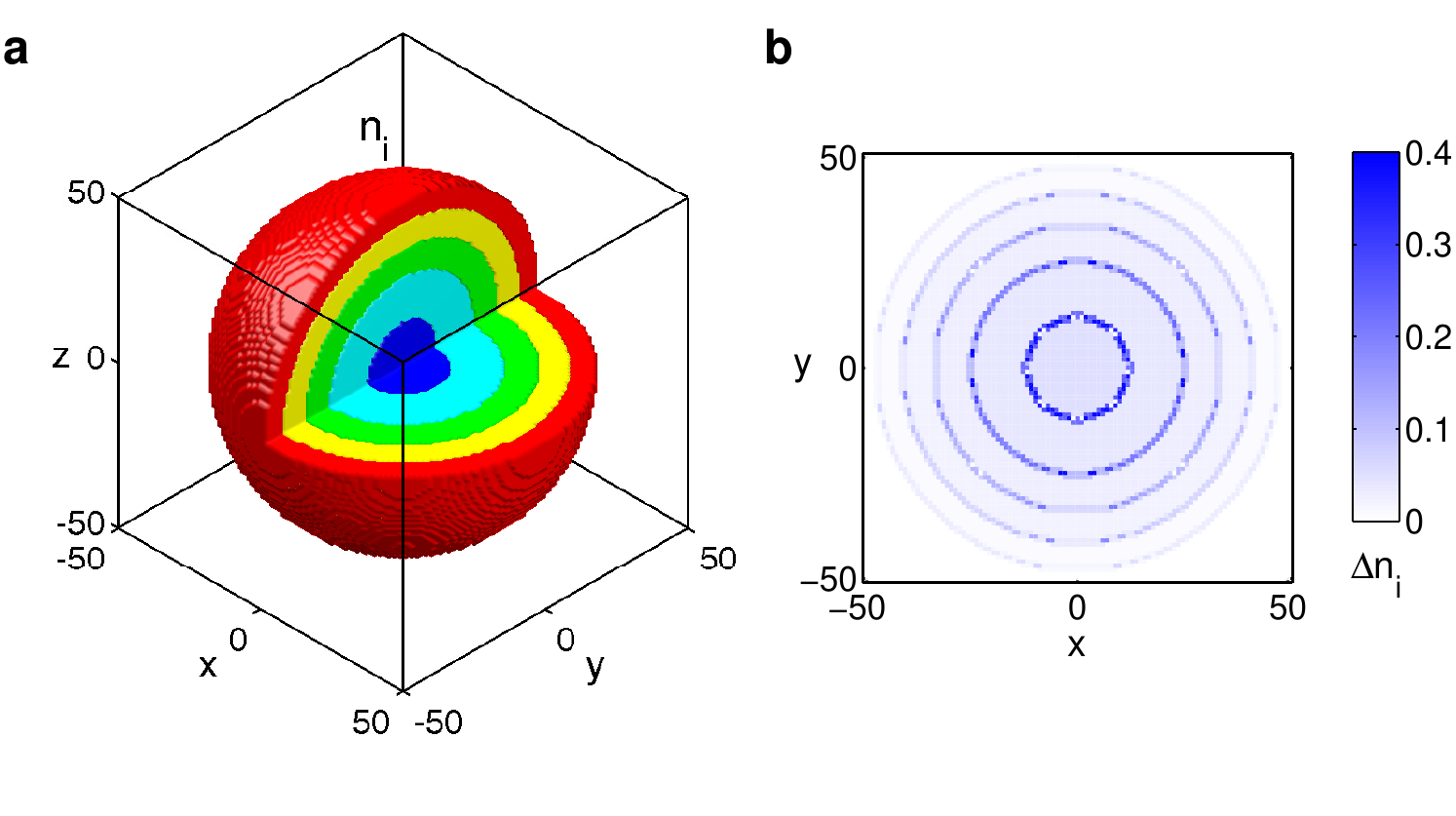}
\end{center}
\vspace{-1cm}
\caption{(a) Cut through the occupation number distribution $n_i$ 
of a 3D lattice ($N=10^6$, $J=0$, $V_h=7.5\times10^{-4}\ER$ and $V_0=25\ER$) 
showing fillings $n_i=1$ (outer shell) to $n_i=5$ (inner shell). 
(b) Particle fluctuations $\Delta n_i$ due to finite tunnelling ($J=10^{-3}\ER\approx 2\times 10^{-3} U$) appear at
the surfaces of the shells ($z=0$ plane).  }
\label{shells}
\end{figure}

Exemplarily, the occupation number distribution $n_i$ for $N=10^6$ particles 
in a 3D lattice is depicted in figure~\ref{shells}a.
It shows the expected shell structure with filling factors five (in the centre)
to one. 
An advantage of the presented numerical algorithm is that 
it is also applicable to 
rapidly varying confinement potentials.  
It can thus be used to tailor more sophisticated shell configurations for experiments. 
Since the on-site interaction energy $U$ is relatively small compared to 
the depth of the lattice wells, the local filling factors can be adjusted without
a stronger perturbation of the 3D lattice.
Additional laser  beams  
or magnetic fields can thus be used to obtain
complex spatial distributions of atoms with specific filling factors.
Figure~\ref{extended}a shows the occupation number distribution for 
an added periodic potential in $x$-direction $V_C=\Vh x^2 / a^2 + U\cos^2(\pi x/a')$
with the periodicity $a'=3a$   
motivated by \cite{Peil:051603}.  
The superlattice structure  leads to
alternating local fillings.
As a second example, an added hat-shaped ($-|x|$) potential
causes two separate atomic clouds,  shown in figure~\ref{extended}b,  that are in touch
with each other at the origin.

\section{Single particle excitations and finite tunnelling}
\begin{figure}
\begin{center}
\includegraphics[width=\widthFullColumn]{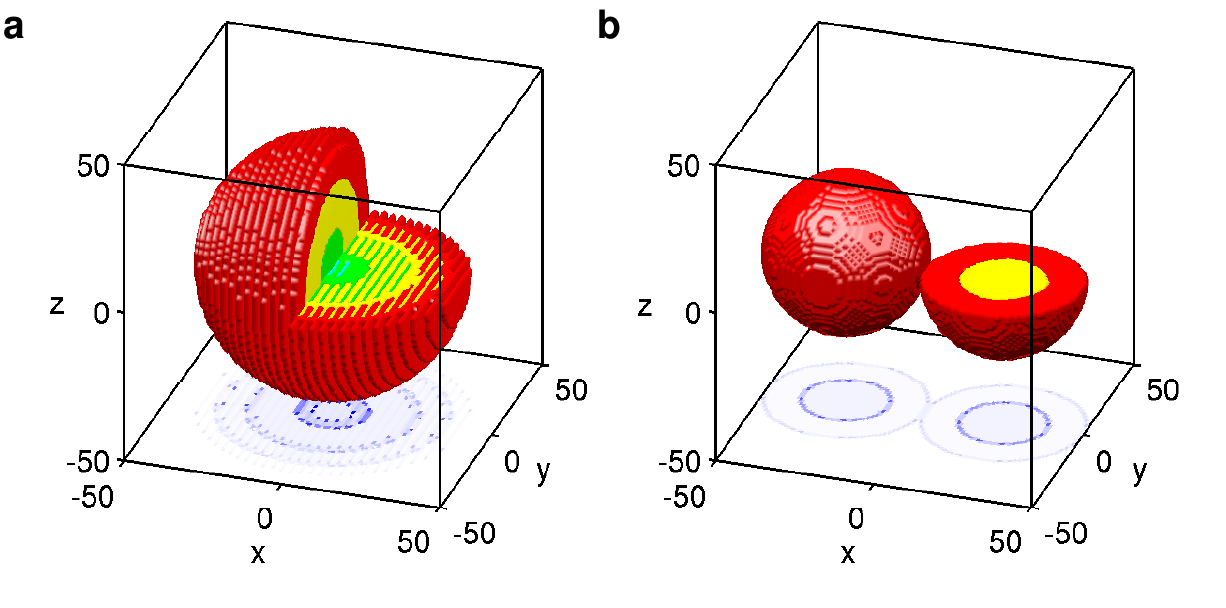}
\end{center}
\vspace{-1cm}
\caption{(a) An additional standing wave with periodicity $3a$ and amplitude $U$ causes alternating fillings in 
$x$ direction ($N=3.5\times10^5$).  
(b) An added $-50\Vh|x|$ potential leads to two separate atom spheres ($N=1.5\times10^5$). 
The fluctuations $\Delta n_i$ for $J=10^{-3}\ER$ and $z=0$ are shown at the bottom 
(see figure~\ref{shells} for the colour map and other parameters).
}
\label{extended}
~\\
\centering
\includegraphics[width=\widthHalfColumn]{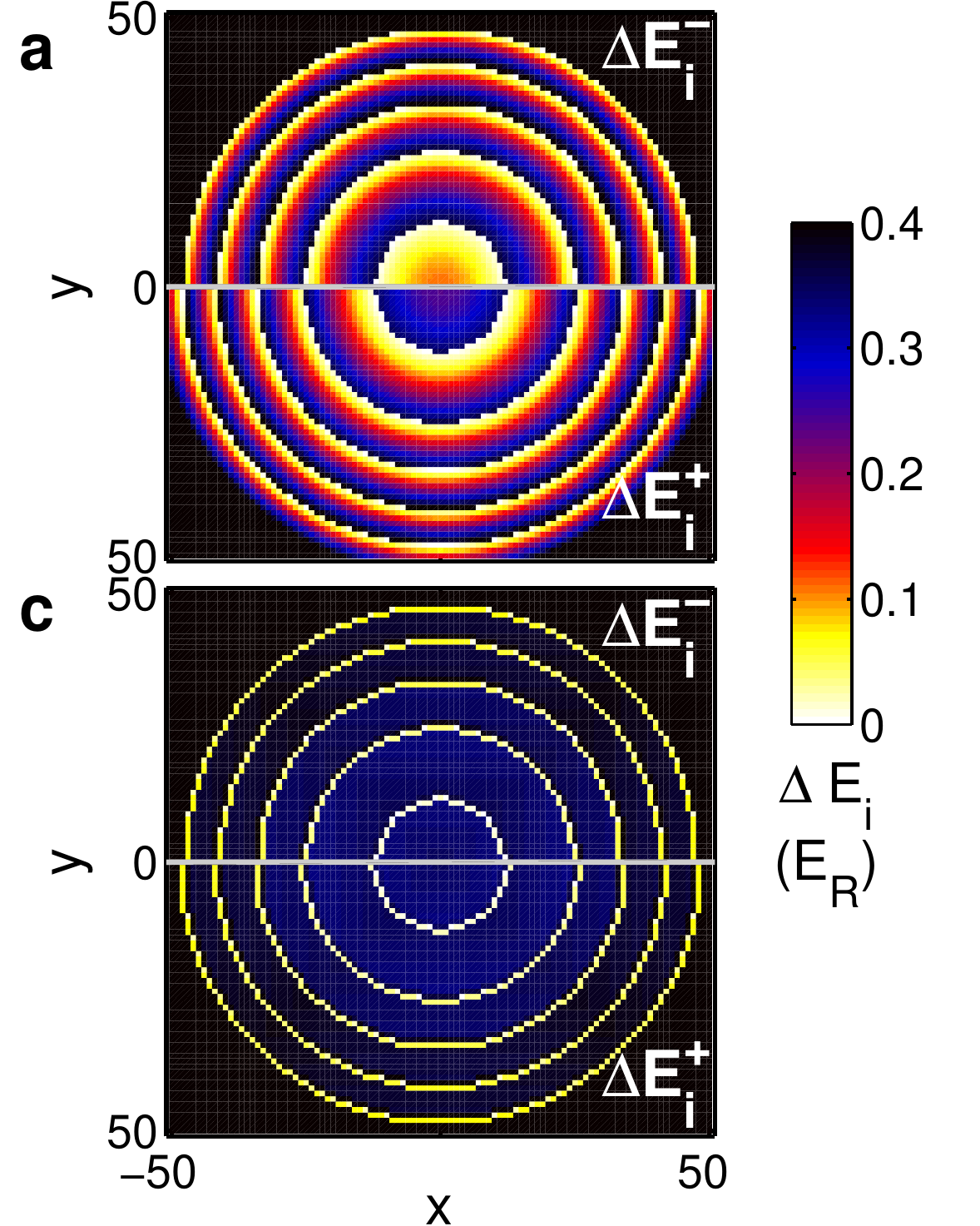}
\includegraphics[width=\widthHalfColumn]{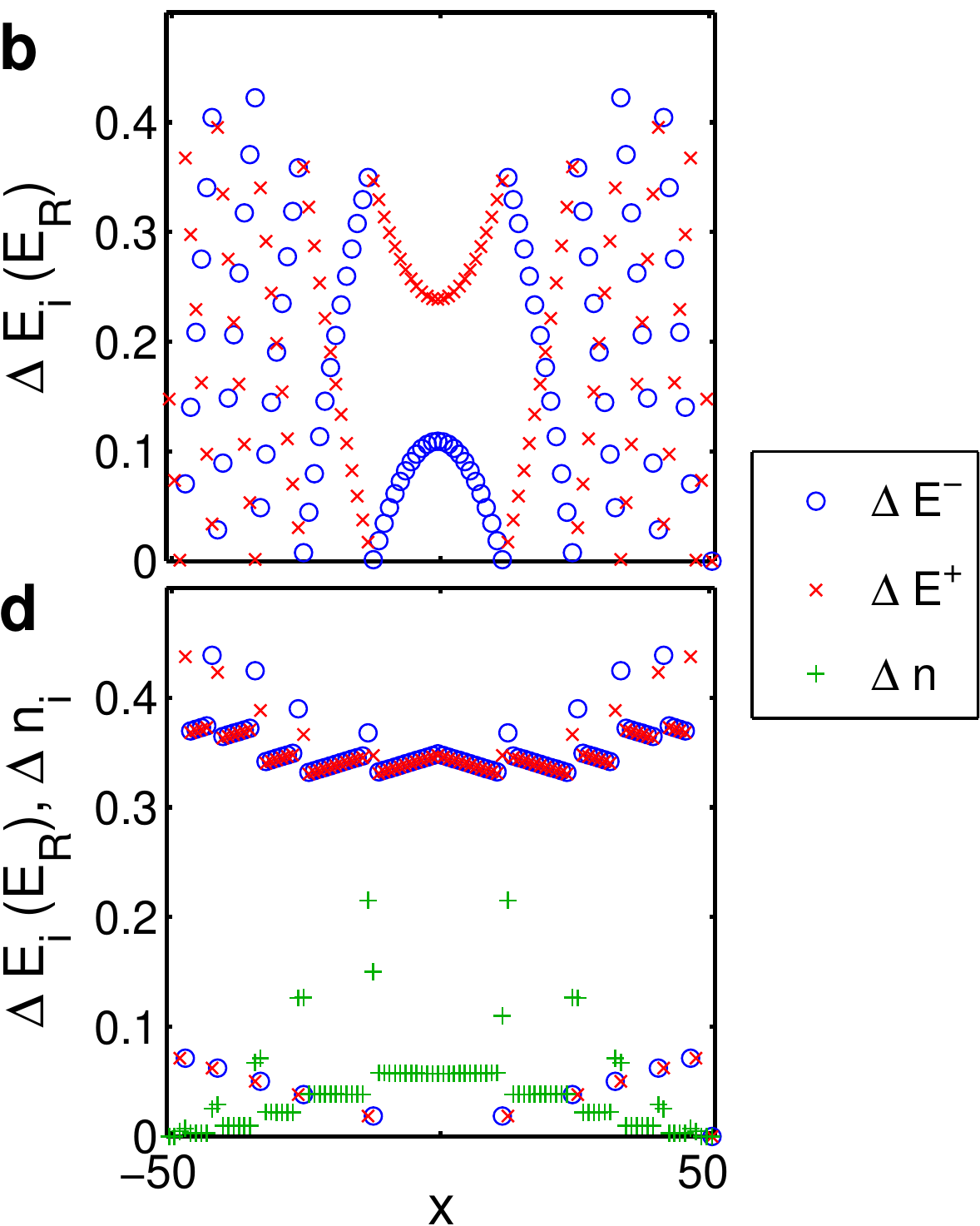}
\caption{The minimal excitation energy in the $z=0$ plane (a) and for $y=z=0$ (b) for particles hopping 
from site $i$ ($E_i^-$) and to site $i$ ($E_i^+$). The excitation energies only accounting for nearest-neighbour hopping in the $z=0$ plane (c) and for $y=z=0$ (d) including particle fluctuations $\Delta n_i$ for $J=5\times10^{-3}\ER$.}
\label{excitations}
\end{figure}

At the outer surface of each shell, almost
gapless excitations are possible via the hopping of a particle to another site, 
whereas the inner surface can easily absorb particles.
This is shown quantitatively in figure~\ref{excitations}a  (upper half),  
where $\Delta E_i^-=\mu_i^- + \min_j( \mu_j^+ )$ is the minimal energy
for removing one particle from site $i$ and adding it to another site $j$.
The minimal energy for adding one particle to site $i$ (and removing it from site $j$)
is denoted as $\Delta E_i^+=\mu_i^+ + \min_j( \mu_j^- )$  (lower half of figure~\ref{excitations}a).  
The local excitation gap  vanishes at the outer and inner surfaces and
increases strongly within the shells.    
The excitation energies  $\Delta E_i^\pm$ are explicitly shown for sites with $y=z=0$ 
in figure~\ref{excitations}b.
It reflects basically the harmonic shape of the confinement, so that  $\Delta E_i^+$ and $\Delta E_i^-$
are positive and negative parables, respectively, subtracted by integer multiples
of the on-site interaction.
The situation changes drastically, if accounting only for excitations to nearest neighbours
as shown in figures~\ref{excitations}c and \ref{excitations}d.  
Low energy excitations  (either $\Delta E_i^+$ or $\Delta E_i^-$)   can occur exclusively on sites
directly at the boundary between different shells.   
Within the bulk the single-particle excitation gap is wide and nearly constant.
Therefore, the nearest-neighbour tunnelling is, in general, strongly suppressed
on these sites, where the system is a good insulator.

In the following, we use a perturbative approach to obtain particle fluctuations
for finite tunnelling $J$. 
For each site $i$ the subsystem containing the site $i$ and all neighbouring
sites is considered. The diagonalization of this subsystem with finite tunnelling 
$-J\sum_{\expect{i,j}}\hat{b}_i^\dagger \hat{b}_j$ allows an approximative calculation of the
particle fluctuations $\Delta n_i=\sqrt{\expect{\hat{n}_i^2}-\expect{\hat{n}_i}^2}$. 
For the example above, weak tunnelling causes finite fluctuations $\Delta n_i$
at the boundaries of the shells,  which is shown in figures~\ref{shells}b and \ref{extended} for $J=10^{-3}\ER$
and in figure~\ref{excitations}d for $J=5 \times 10^{-3}\ER$.   
Please note, that due to the finite cell size of the lattice, the results are not
completely spherically symmetric.  
In accordance with the nearest-neighbour excitation energies, the fluctuations affect only
sites directly at the surfaces.
Because of the larger slope of the confinement and
the lower filling, the tunnelling decreases for outer shells.
Within the bulk of the Mott shells small particle fluctuations can be observed due to the finite tunnelling.
The fluctuations on the surface however enhance the tunnelling of atoms next to the surface, which
is not covered by this  perturbative approach   
and would require a self-consistent calculation.
The narrow energy gaps of particles close to the surface can, in principle, cause
near-resonant tunnelling to non-nearest neighbours, so-called variable range hopping \cite{Ambegaokar:2612}.
Due to the relatively strong $\Vh$ (compared with $J$) and the regularity of the 
potential, only sites on the surface, which are dominated by nearest-neighbour hopping, 
have suitably low excitation energies.

The presented method provides the lowest energetic states 
(including particle-hole excitations) and is suitable to restrict the Hilbert space effectively.
The obtained states can be used as a starting point for  
exact diagonalization, quantum Monte-Carlo, and dynamical mean-field calculations including 
tunnelling and finite temperatures.  
In particular, the diagonalization of the 1D lattice (figure~\ref{spectrum}b)  
could be transformed with much fewer basis states.

\section{Conclusions}
In summary, we have studied the excitation spectrum and exact site occupation numbers  
for confined optical lattices deep within the Mott insulator regime.
In good agreement with exact diagonalization for small 1D lattices, a numerical method
was presented that allows  for negligible  tunnelling 
the exact treatment of macroscopic optical lattices with arbitrary shape of the confining potential.  
Adding slowly varying potentials to the optical lattice can give rise to 
complex filling structures.   
We have calculated the numerically exact many-particle on-site energies and have shown  
that introducing a filling factor depending on-site interaction can be 
incorporated to cover orbital changes.  
For small systems, the many-particle spectrum contains lobes,  
whereas for macroscopic systems
nearly gapless excitations are always possible 
at the boundaries of the Mott shells.  
Within a given Mott shell the local excitation energy varies strongly
leading to compressible sites close to the surface and incompressible inner sites.
A perturbative treatment for finite tunnelling shows strong particle fluctuations at the boundaries
between the shells, where the Mott-insulator gap vanishes.  
Finally, the presented method can serve as a reduction scheme for the Hilbert space 
for further numerical treatments.  

\section{Acknowledgments}
We would like to thank P.~Ernst and K.~Sengstock for valuable discussions.


\begin{thebibliography}{18}
\bibitem[16]{note2}
\bibinfo{note}{For certain system parameters ($N$, $m_x$, $\Vh$) a degeneracy
  of the ground state can occur.}
\end{thebibliography}
\end{document}